\pdfoutput=1
\documentclass[fleqn,usenatbib]{mnras}

\usepackage{newtxtext,newtxmath}
\usepackage{aas_macros}
\usepackage{hyperref}
\usepackage{natbib}
\usepackage[T1]{fontenc}
\usepackage{graphicx}
\usepackage{orcidlink}
\usepackage{placeins}
\DeclareRobustCommand{\VAN}[3]{#2}
\let\VANthebibliography\thebibliography
\def\thebibliography{\DeclareRobustCommand{\VAN}[3]{##3}\VANthebibliography}

\usepackage{graphicx}	
\usepackage{amsmath}	

\title[The nature of ASASSN-24fw's occultation]{The nature of ASASSN-24fw's occultation: modelling the event as dimming by optically thick rings around a sub-stellar companion\thanks{This paper includes data gathered with the 6.5 m Magellan Clay Telescope at Las Campanas Observatory, Chile.}}

\author[S. Shah et al.]{
Sarang Shah$^{1,13}$, \orcidlink{https://orcid.org/0000-0003-1959-8439}\thanks{E-mail: sshah1502@gmail.com},
Jonathan P. Marshall$^{2}$ \orcidlink{https://orcid.org/0000-0001-6208-1801},
Carlos del Burgo$^{3,4}$ \orcidlink{https://orcid.org/0000-0002-8949-5200},
Gergely Hajdu$^{1}$ \orcidlink{https://orcid.org/0000-0003-0594-9138},
Isabel Rebollido$^{5}$ \orcidlink{https://orcid.org/0000-0002-4388-6417}, \newauthor
Bogumił Pilecki$^{1}$ \orcidlink{https://orcid.org/0000-0003-3861-8124},
Ashish Mahabal$^{6}$ \orcidlink{https://orcid.org/0000-0003-2242-0244},
Sascha T. Zeegers $^{8}$ \orcidlink{https://orcid.org/0000-0002-8163-8852},
Bacham Eswar Reddy$^{9}$ \orcidlink{https://orcid.org/0000-0001-9246-9743}, \newauthor
Franciska Kemper$^{10,11,12}$ \orcidlink{https://orcid.org/0000-0003-2743-8240},
Mansi M. Kasliwal $^{6}$ \orcidlink{https://orcid.org/0000-0002-5619-4938},
Viraj Karambelkar $^{6}$\orcidlink{https://orcid.org/0000-0003-2758-159X},
Matthew J. Graham $^{6}$ \orcidlink{https://orcid.org/0000-0002-3168-0139}, \newauthor
S. G. Djorgovski $^{6}$ \orcidlink{https://orcid.org/0000-0002-0603-3087}, 
Daniel Stern $^{7}$ \orcidlink{https://orcid.org/0000-0003-2686-9241}
\\
$^{1}$Nicolaus Copernicus Astronomical Center, Polish Academy of Sciences, Bartycka 18, 00-716 Warsaw, Poland\\
$^{2}$Institute of Astronomy and Astrophysics, Academia Sinica, 11F of AS/NTU
Astronomy-Mathematics Building, No.1, Sec. 4, Roosevelt Rd, Taipei 106319, Taiwan\\
$^{3}$Instituto de Astrof\'isica de Canarias, Vía L\'actea S/N, La Laguna, E-38200, Tenerife, Spain\\
$^{4}$Departamento de Astrof\'isica, Universidad de La Laguna, La Laguna, E-38200, Tenerife, Spain\\
$^{5}$ European Space Agency (ESA), European Research and Technology Centre (ESTC), Camio Bajo del Castillo s/n, 28692 Villanueva de la Ca$\~{n}$ada, Madird Spain\\
$^{6}$Division of Physics, Mathematics and Astronomy, California Institute of Technology, Pasadena, CA 91125, USA\\
$^{7}$ Jet Propulsion Laboratory, California Institute of Technology, 4800 Oak Grove Drive, Pasadena, CA 91109, USA\\
$^{8}$ European Space Agency (ESA), European Research and Technology Centre (ESTC), Keplerlaan 1, 2201 AZ Noordwijk, The Netherlands\\
$^{9}$ Indian Institute of Technology, NH-44 , PO Nagrota, Jagti, Jammu and Kashmir 181221, India\\
$^{10}$ Institut de Ciències de l'Espai (ICE, CSIC), Can Magrans, s/n, E-08193 Cerdanyola del Vallès, Barcelona, Spain\\
$^{11}$ ICREA, Pg. Lluís Companys 23, E-08010 Barcelona, Spain\\
$^{12}$ Institut d'Estudis Espacials de Catalunya (IEEC), E-08860 Castelldefels, Barcelona, Spain\\
$^{13}$ Inter-University Centre for Astronomy and Astrophysics, Ganeshkhind, Pune, Maharashtra, India - 411007\\
}

\date{Accepted XXX. Received YYY; in original form ZZZ}

\pubyear{\the\year{}}

\begin{document}
\label{firstpage}
\pagerange{\pageref{firstpage}--\pageref{lastpage}}
\maketitle

\begin{abstract}
ASASSN-24fw, monitored by the All-Sky Automated Survey for SuperNovae, underwent a rapid and deep dimming event beginning in late 2024 and lasting until June 2025. The pre-dimming spectral energy distribution indicates that ASASSN-24fw is an F-type main-sequence star with a persistent infrared excess, corresponding to a fractional luminosity of about 10 per cent. Public survey lightcurves show that the star is otherwise photometrically stable, implying that the dimming was caused by an external occulter. While long-duration stellar dimming events have become increasingly common in recent years, ASASSN-24fw is distinguished by a pronounced flat-bottomed lightcurve lasting nearly 200 days. We analyse this event using available photometric and spectroscopic data obtained during the dimming phase. Two independent lightcurve models are applied. The first characterizes the multiple phases of the ingress structure, while the second model suggests that the occulter is a gas-giant companion, likely a brown dwarf, with a minimum mass of 3.42 M\textsubscript{J} and an extended ring system with a radial extent of approximately 0.17 au. Near-infrared spectra obtained during dimming show enhanced infrared excess and are consistent with an M8-type brown dwarf companion. Optical spectra reveal variable H\textsubscript{$\alpha$} emission during the event. Post-dimming spectra show no evidence for ongoing accretion, suggesting that the H\textsubscript{$\alpha$} variability is associated with the occulter rather than the host star. We conclude that ASASSN-24fw represents a rare and unusual system, and that further observations are required to better constrain its stellar properties, circumstellar environment, and evolutionary history.
\end{abstract}




\section{Introduction}\label{section:introduction}
Nearly all stars are born surrounded by dense circumstellar envelopes \citep{ribas2014,manara2023}. These envelopes process the natal proto-stellar cloud into planetary discs over a period of a few to 10s of million years \citep{williams2011}. These circumstellar young protoplanetary discs are the birthplace of planetary systems including planetesimal belts \citep{wyatt2015, lovell2021}.  
circumstellar discs manifest in various forms \citep{hughes2018} and now we know that they are ubiquitous throughout the pre- to post-main sequence lifetimes of stars \citep{jewitt2009, dennihy2020, zhao2020,  raymond2022, myers2024, swan2024, farihi2025}. The dynamical excitation induced erosive or destructive collisions of the planetesimals (e.g. the proposed \textit{Late Heavy Bombardment} period in the Solar System \cite{gomes2005, bottke2017}) is a common mechanism to generate detectable levels of dust manifesting as debris discs around stars \citep[e.g.][]{marshall2021, matra2025}.

 Circumstellar discs significantly change their structure, dynamics, and composition over time and, thus their architectures and brightness are correlated with the stellar age \citep{cao2023, matra2025}, e.g., a pre-main sequence is surrounded by a protoplanetary disc of mainly gas and dust, the transition or intermediate disc phase has altered composition and geometry while the debris disc is further evolved like the star hosting it and contains mainly of collisional rocky bodies and dust generated by destructive processes \citep{hughes2018, su2019, moor2024}. Thus, the detection of such debris discs points towards successful planet-formation (or even destruction) processes.

These circumstellar discs can be detected through the presence of excess flux from infrared to millimeter wavelengths \citep{su2019}. As the material in these discs re-emits the incident stellar radiation into longer wavelengths, the composition of the debris disc, structure and its evolutionary status can be inferred from the location of the peak of the excess flux \citep{wyatt2015,hughes2018,cao2023}. This information in parallel with the nature of the optical variability can be used to understand the nature of the circumstellar material e.g., either proto-planetary, transitional, or debris discs. For example, V488 Per has the most infrared excess of any circumstellar disc and is therefore thought to be surrounded by an {\it Extreme Debris Disc} \citep{moor2024} which has a fractional luminosity $>$1\%, HD 166191 is surrounded possibly by a transition disc \citep{kennedy2014} although this is still controversial \citep[see e.g.][]{garcia2019}, and RZ Piscium contains a debris disc perturbed by a red-dwarf companion \citep{kennedy2020}. However, there are examples of optical variability without infrared excess e.g. Boyajian's star \citep[KIC 8452642;][]{boyajian2016} had no transient infrared excess \citep{marengo2015} and therefore it was attributed to the disintegration of a family of exo-comets \citep{bodman2016}. Another such event around ASASSN-21qj has likewise been attribted to exocometary activity and exhibited a transient infrared excess \citep{marshall2023}.

Sometimes, the variability in stellar brightness is also caused by systems containing circumplanetary, or circumstellar discs. For example, Mamajek's object \citep{mamajek2012}, RW Aur \citep{rodriguez2013, rodriguez2016}, V409 Tau and AA Tau \citep{rodriguez2015}, "The Random Transitor" type objects HD 139139 \citep{rappaport2019}, and EPIC 204376071 \citep{rappaport2019b}, Gaia21bcv \citep{hodapp2024}, ASASSN-21js \citep{pramono2024}. Another example is VVV-WIT-08/10 \citep{smith2021}, discovered by the VISTA Variables in the V\'{i}a-L\'{a}ctea \citep[VVV;][]{minniti2010} survey, where an eclipse lasting $\simeq$ 200 days is attributed to an elliptical disc-like occulter, similar to the proposed ring system of J1407b \citep{kenworthy2015}.

One of the important distinguishing points between variability due to circumstellar material and circumplanetary material is the depth and duration of the dimming. While the variability due to the former is irregular and a combination of several short-period dips, the variability due to the latter could manifest as a dip both larger in depth and period. It may also show some extra features depending on the orientation of the system. As it is not exactly clear how such objects with  optically thick rings form and if there is any correlation between the circumstellar environment and the properties of such objects, we can build a clearer picture about the such systems only by continuing to identify and analyze such events. 

This paper is another attempt to explain the recent stellar dimming event: ASASSN-24fw. This dimming event was reported by ASAS-SN's transient detection page followed by an Astronome$g^{\prime}$s Telegram \citep[ATel No. 16833,][]{johantgen2024}. ASASSN-24fw has been the subject of study in two recently published articles by \cite{toribio2025} and \cite{zakamska2025}. 
In September 2024, it experienced a steep reduction in brightness which lasted until around the first week of October 2024. The star remained around the same low-brightness until the mid-May of 2025 and the egress occurred close to the first week on June 2025. Another Telegram by \citet{nair2024} (ATel No. 16919) noted similar dimming episodes of this star around the years 1937 and 1982, suggesting a possible periodicity of $\simeq$ 16,000 days (or 43.8 years). The drastic drop in the brightness of ASASSN-24fw from $g^{\prime}$ $\simeq$ 13.20 mag to $\simeq$ 16.70 mag (close to the limit of ASAS-SN sensitivity at $\sim$ 17.50 mag) rapidly within a fortnight followed by a nearly stable brightness for six months and mirrored egress is the unique feature of this event.

\section{Data}\label{section:data}
\subsection{All-Sky Automated Survey for SuperNovae (ASAS-SN)}\label{section:asassn}
The All-Sky Automated Survey for SuperNovae (ASAS-SN) \citep{shappee2014} is a long term survey project to monitor the sky down to a limiting magnitude of V $\simeq$ 17 mag using a global network of telescopes. The focus of the survey is to find nearby supernovae (SNe) and other transient sources. Currently, this network consists of six fully robotic units on Mount Haleakala in Hawaii, Cerro Tololo in Chile, LCO sites in South Africa and USA, and another in China. Each unit consists of several robotic 14 cm aperture telescopes and is hosted by the corresponding station and each telescope is mounted with a Nikon telephoto lens and a 2k x 2k thinned CCD, giving a 4$\fdg$5$\times$4$\fdg$5 field-of-view and a 7$\farcs$8 pixel scale. These 24 telescopes together survey entire night sky every night and discoveries are announced within few hours of the data being taken. The data can be accessed through \url{https://asas-sn.osu.edu/}.

\subsection{LCOGT}\label{section:lcogt}
We monitored ASASSN-24fw with time-series photometry from February 22\textsuperscript{nd} to April 5\textsuperscript{th} 2025 (PID:DDT2025A-003, PI: Sarang Shah) using the Las Cumbres Observatory's network of 1-m telescopes with the SINISTRO imaging instrument \citep{brown2013}. The stellar brightness was recorded in four filter bands: SDSS $g^{\prime}$, $r^{\prime}$, $i^{\prime}$, and PANSTARRS z\textsubscript{s}. Integration times were 12s in $g^{\prime}$, 18s in $r^{\prime}$, 26s in $i^{\prime}$, and 94s in z\textsubscript{s}. This provided a SNR$\geq$100 in each filter for an occulted magnitude of $\simeq$ 16.5. The photometric measurements were taken from the pipeline-produced BANZAI catalog \citep{mccully2018}. If the catalog (or target) was missing from the observatory provided data, the stellar magnitude was determined using aperture photometry performed using the Python-based {\sc Photutils} package \citep{bradley2024} using a 5$\arcsec$ radius aperture and a background annulus between 10$\arcsec$ and 15$\arcsec$.

\begin{figure}
    \centering
    \includegraphics[width=\columnwidth]{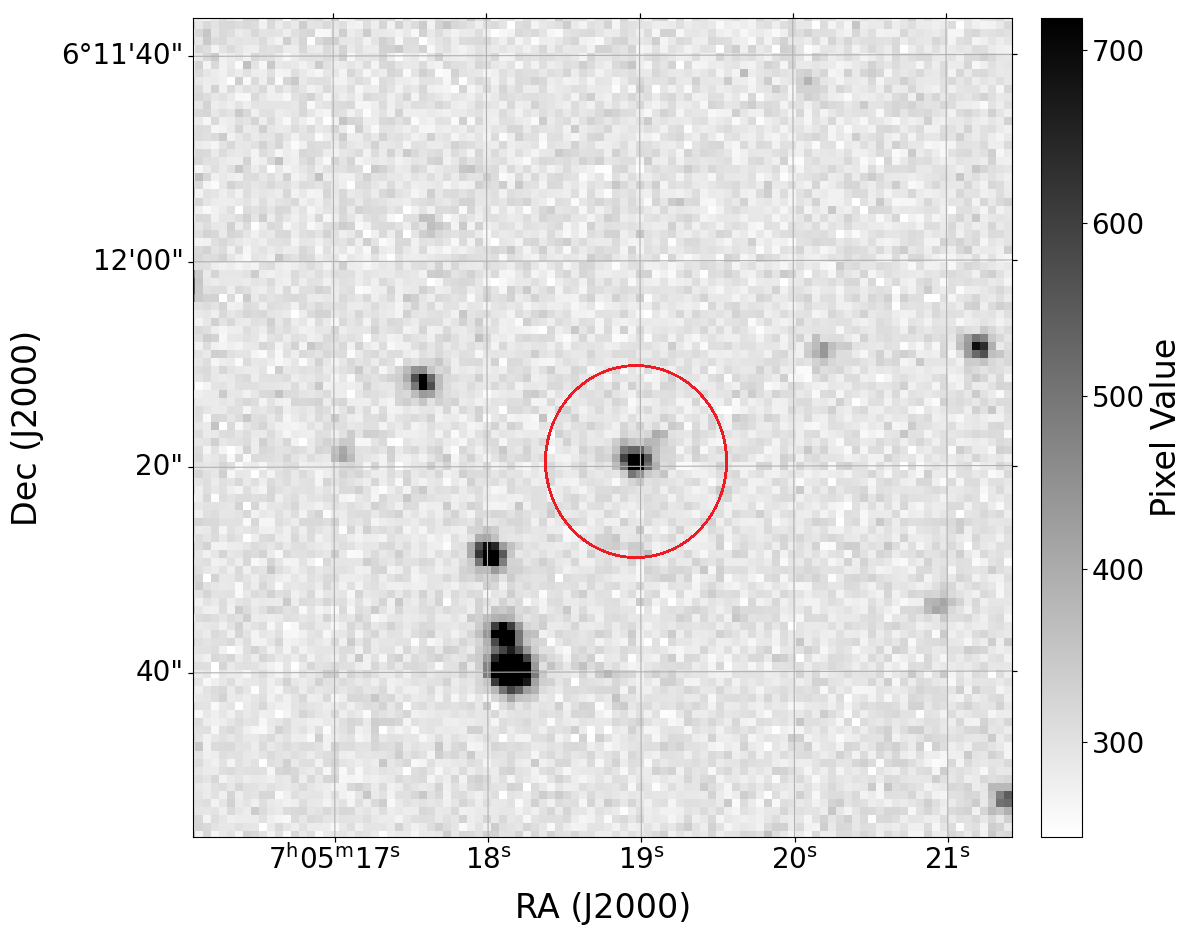}
    \caption{LCOGT g-band post-dimming image of ASASSN-24fw in the center marked by an orange colored circle. The colorbar indicates the pixel value. We can also see a fainter companion inside the same circle which is at 3$\arcsec$ separation in Gaia DR3. At a distance of $\simeq$ 1kpc, this companion is separated by $\simeq$ 3000 au.}
    \label{fig:lcogtASASSN-24fw}
\end{figure}

In the LCOGT images, we find a faint companion star which is also present in \textit{Gaia} DR3 with $G$ > 19 mag and at a separation of 3$\arcsec$. In Figure (\ref{fig:lcogtASASSN-24fw}), we show a cropped 3$\arcsec\times$3$\arcsec$ LCOGT $g^{\prime}$-band image where the colorbar represents the pixel count. ASASSN-24fw and its faint companion are located in the center of the red-circle (and image). If bound to ASASSN-24fw, at a distance of $\simeq$ 1kpc, this companion is located at a distance of 3000 au. To check how likely it is to have a chance alignment for a faint companion at a separation 3$\arcsec$, we stacked our images to increase the SNR of the companion. We then performed the psf photometry of the companion by masking ASASSN-24fw. The ($g^{\prime}$-$r^{\prime}$), ($r^{\prime}$-$i^{\prime}$), and ($i^{\prime}$-$z^{\prime}$) colors that we get are 1.0, 0.66, and -0.83 respectively. The former two colors indicate a K-type main-sequence star. The negative ($i$-$z$) color is due to bad sky in the $z$-band images. Further, we find there are 439 sources fainter than or equal to the brightness of ASASSN-24fw. We define the chance alignment probability $P$($x$) of a star within 3$\arcsec$ radius of ASASSN-24fw as:

\begin{equation}
\centering
P  \left( x \right) = \frac{A_{3}\times N_{\rm Sources}}{FoV}
\end{equation}

where the FoV is 13.2 $\times$ 13.2 sq. arcmin and $A_{3}$ is the area of a circular region of 3$\arcsec$ radius around ASASSN-24fw. We find that $P$($x$) is only 2\% which indicates that there is a high chance that the companion might not be a chance alignment. But this can be only confirmed in future by conducting follow-up campaign to study the proper motion of this faint companion.
\subsection{AAVSO}\label{section:aavso}
American Association of Variable Star Observers (AAVSO) \citep{kloppenborg2025} is an association of professional and amateur astronomers that observe variability related phenomena and catalog their high quality photometric observations. This association of professional and amateur astronomers regularly observes and analyses variable stars and their observers are located across the globe. We obtained the AAVSO observed data from their website\footnote{\url{https://www.aavso.org/data-download/}}. 

\subsection{NEOWISE}\label{section:neowise}
The NEOWISE mission \citep{neowise2020} utilized the \textit{Wide-field Infrared Survey Explorer} (WISE) spacecraft and a survey strategy similar to WISE mission \citep{wright2010} to scan the sky between the ecliptic poles in great semi-circles and nearly perpendicular to the Earth-Sun line. This survey strategy created a set of multiple independent exposures on each point of the sky over the 10 year period and had observed the region including ASASSN-24fw in the past decade. Thus, we could  obtain and combine all the single exposure NEOWISE photometry which is an array of one observation every six months. We found no infrared variability of ASASSN-24fw in the NEOWISE lightcurve as all NEOWISE observations were taken outside the eclipse.

\subsection{TESS}\label{section:tess}
The Transiting Exoplanet Survey Satellite (TESS) is a satellite designed to search for transiting exoplanets among the brightest (and nearest) stars over most of the sky \citep{ricker2015}. TESS orbits the Earth every 13.7 days on a highly elliptical orbit, scanning a sector of the sky spanning 24 x 96 deg\textsuperscript{2} for a total of two orbits, before moving on to the next sector. It captures images at cadences of 2 s (used for guiding), 20 s (for 1\,000 bright astroseismology targets), 120 s (for 200\,000) and 30 min (full frame images). The instrument consists of 4 CCDs each with a field of view of 24 x 24 deg\textsuperscript{2}, with a wide band-pass filter from 600-1000 nm.

ASASSN-24fw has a TESS ID (TIC) = 262032775. It was detected in Sectors 7 and 33, with observations taken from 8 January 2019 to 1 February 2019 and from 17 December 2020 to 13 January 2021. This was the period when the star was bright and therefore we could obtain its lightcurve to search for periodicity. We did not find any peaks in the periodogram, suggesting that ASASSN-24fw is a slow rotator and has no significant intrinsic variability. Although ASASSN-24fw was again observed by TESS in December 2024, there is no lightcurve available due to its faintness.

\subsection{Spectral Observations}\label{section:spectralobs}
We obtained a low resolution spectrum with Keck LRIS instrument \citep{oke1995, rockosi2010} on 29 October 2024. The resolution of the grating element was $\sim$ 1300 and the exposure time was 600s. We also obtained a second high resolution transmission spectrum using the HiRES instrument on 28 January 2025. The resolution of the grating element that was used was $\sim$ 25,000 and the exposure time was 2700s. Both of these observations were conducted when the star was dimmed and were extracted using \texttt{Pypelt} \citep{prochaska2020}. The wavelengths were then calibrated using a combination of Th-Ar spectra and telluric absorption with the standard star observations using \texttt{xtellcor} \citep{vacca2003}. We also obtained a single NIR low resolution spectrum (R$\simeq$ 3000) using the TripleSpec instrument \citep{herter2008} at the 5.1m Hale Telescope at Palomar Observatory on 21 December 2024. The exposure time was 120s and the spectrum was extracted using the IDL package \texttt{spextool} \citep{cushing2004}. The extracted spectrum was also flux calibrated and corrected for telluric absorption in a similar way. We recently obtained two post-dimming optical spectra of ASASSN-24fw on 13 October 2025 using the MIKE spectrograph on Magellan Telescope. The exposure time of each spectrum was 600s with a effective resolution of 57,000 and they were reduced using Daniel Kelson’s pipeline available at the Carnegie Observatories Software Repository\footnote{http://code.obs.carnegiescience.edu/}. This was taken to compare the during-dimming and out-of dimming nature of the star. We show this raw stacked spectrum in Figure (\ref{appendixfig:a}) of Appendix (\ref{appendix:a}).

\subsection{Spectral Energy Distribution}\label{section:sed}
\begin{figure*}
    \centering
\includegraphics[height=100mm,width=0.95\textwidth]{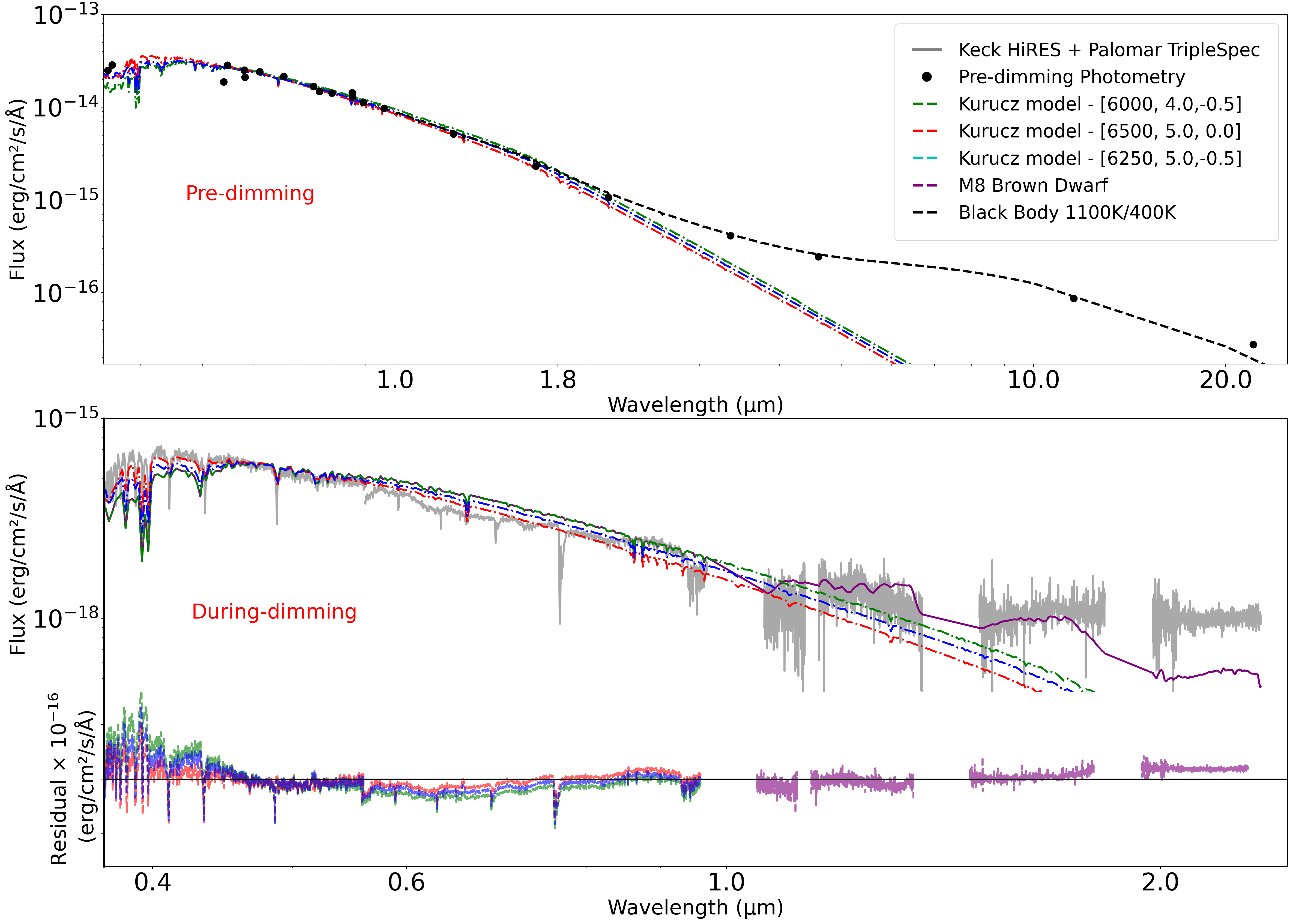}
\caption{ {\it Top}: Pre-dimming SED of ASASSN-24fw formed using various optical to infrared surveys photometry. The infrared excess can be seen towards the longer mid-infrared wavelengths which can be fitted by a two component blackbody model. \textit{Bottom}: Keck HiRES and Palomar TripleSPEC spectra that we obtained during-dimming. For reference, we have also plotted the M8 brown dwarf template in purple color, the continuum of which matches very well to the initial part of the observed NIR spectrum. The excess flux in the latter part of the NIR spectrum hints at elevated levels of infrared excess perhaps due to circumplanetary rings to the brown dwarf that we discuss in this paper. The panel also shows the residual where the three different Kurucz model spectra and the template brown dwarf spectrum are subtracted from the observed optical and NIR spectra.}
\label{fig:sed0}
\end{figure*}

We form the spectral energy distribution of ASASSN-24fw using publicly available pre-dimming photometry from various surveys. For this, we obtained single epoch photometric data in the optical from SDSS \citep{almeida2023}, PANSTARRS DR2 \citep{flewling2018},  \textit{Gaia} DR3 \citep{gaia2023}, \textit{HST} \citep{whitmore2016}, and in the infrared 2MASS \citep{skrutskie2006} and \textit{WISE} \citep{wright2010}. We corrected these magnitudes (as well as our observed spectra during dimming) for extinction using the mean value of pre-dimming reddening from \cite{schlafly2011} reddening maps. These maps are updated from \cite{schlegel1998} reddening maps and use the standard \cite{fitzpatrick1999} reddening law. We find a relatively low value of E(B-V) $\simeq$ 0.05 for ASASSN-24fw and a clear presence of infrared excess (around 10\%) in the WISE points as shown in the first panel of Figure (\ref{fig:sed0}). In the second panel of the Figure (\ref{fig:sed0}), we can also see the SED formed using the spectrum observed during the dimming. In the following sections, we describe in detail about our SED fitting  methods. 
\begin{figure*}
\centering
\includegraphics[width=1.0\textwidth]{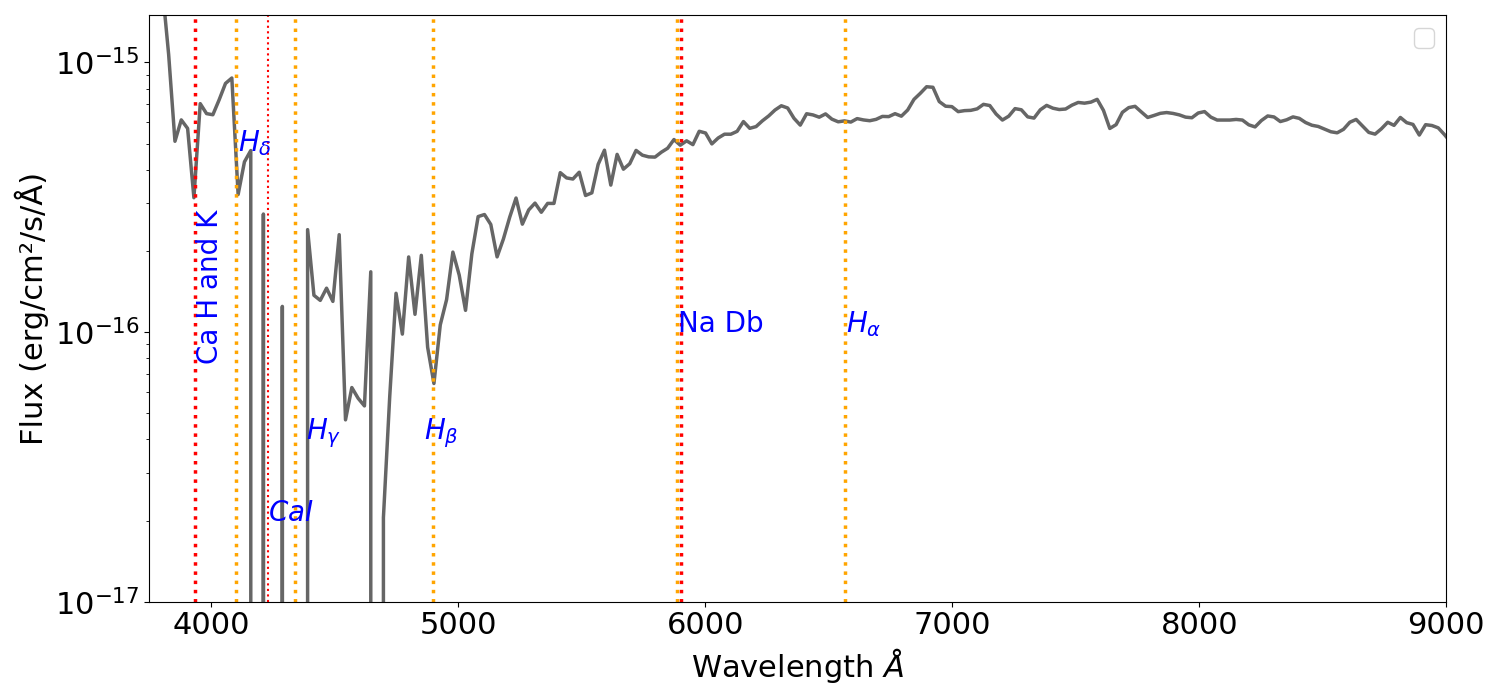}
\includegraphics[width=1.0\textwidth]{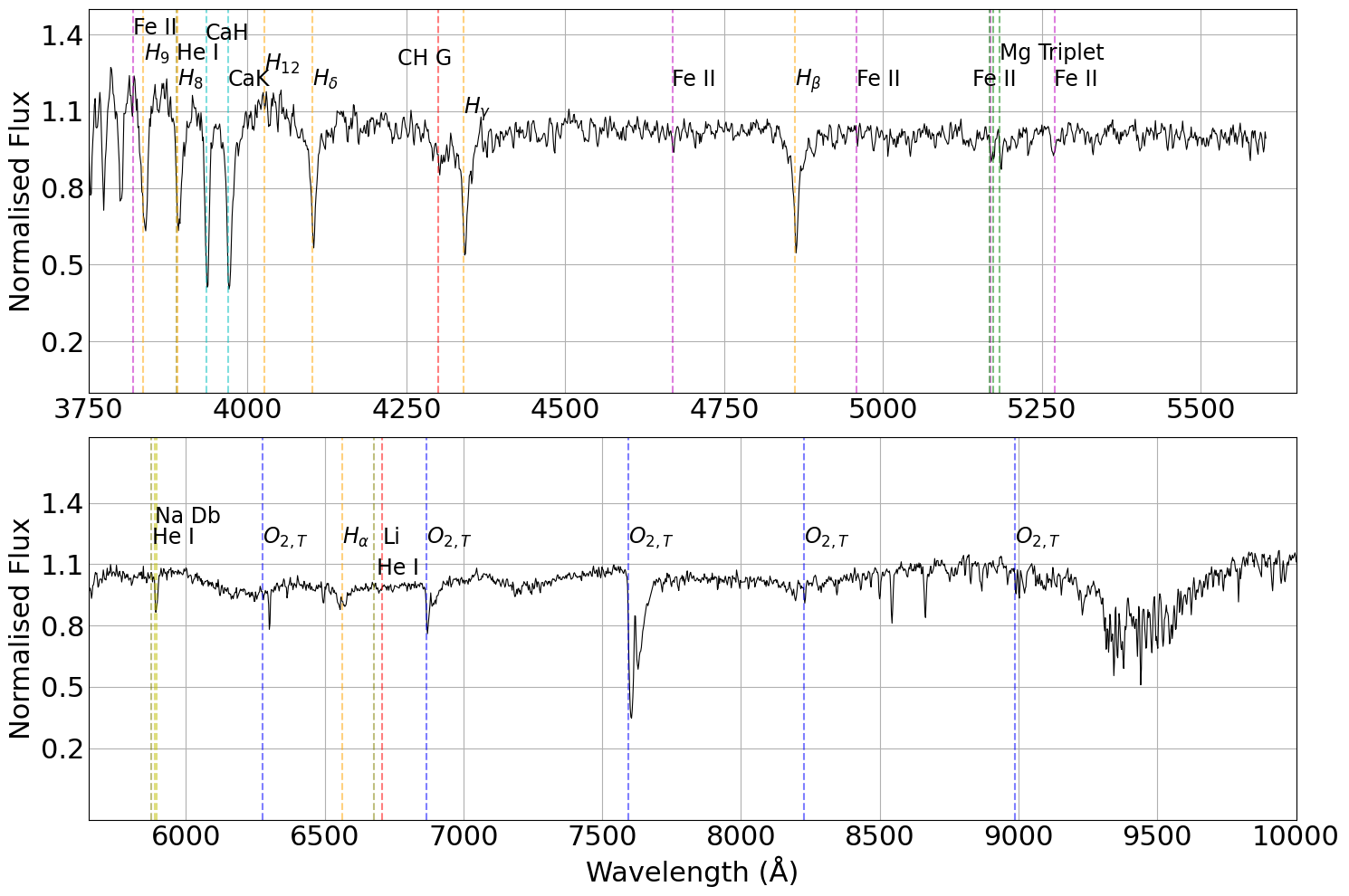}
\caption{\textit{Top}: Low resolution spectrum taken on 29\textsuperscript{th} October 2024 where the location of Hydrogen Balmer series, Ca II H\&K as well as Na Db lines are shown. \textit{Bottom}: High-resolution and high SNR spectrum on 25\textsuperscript{th} January 2025. Here Ca II H and K lines appear to be deeper than the hydrogen Balmer series. The H$\alpha$ emission line is missing from both the spectra, although we see a central emission {\it w} shaped feature in the H$\alpha$ absorption. Further, we see a weak CH G-band molecular line which is prominent in G-type and cooler stars.}
\label{fig:lines}
\end{figure*}
\section{Nature of ASASSN-24fw}\label{section:properties}
\begin{table}
\centering
\caption{Gaia DR3 information about ASASSN-24fw. The proper motion is also converted to 3D peculiar velocity ($U_{s},V_{s},W_{s}$) in Milky Way. The stellar atmospheric parameters provided by \textit{Gaia} DR3 are after modeling the low resolution pre-dimming \textit{Gaia}BP/RP (XP) spectrum.}
\label{table:gaia}
\begin{tabular}{ |c|c| } 
 \hline
 Parameter & Value \\
 \hline\hline
 Gaia ID & \textit{Gaia} DR3 3152916838954800512 \\
 ALLWISE ID & J070518.97+061219.4\\
 Parallax (mas)& 0.9583 $\pm$ 0.0152\\ 
 $\mu_{\alpha}^{\star}$ (mas/yr) & -3.75 $\pm$ 0.01 \\
 $\mu_{\delta}$ (mas/yr) & -7.61 $\pm$ 0.01 \\
 $\mu_{l}$ (mas/yr) & 5.13 $\pm$ 0.01\\
 $\mu_{b}$ (mas/yr) & -6.77 $\pm$ 0.02\\
  Radial Velocity (km/s)& 36.87 $\pm$ 3.60\\
 $v_{l}$ (km/s) & 5.33 $\pm$ 0.13\\
 $v_{b}$ (km/s) & -7.06 $\pm$ 0.28 \\
 $U_{s}$ (km/s) & -10.04 $\pm$ 0.44 \\
 $V_{s}$ (km/s) & 204.04 $\pm$ 0.05\\
 $W_{s}$ (km/s) & -21.70 $\pm$ 0.92\\
 ruwe & 0.97 \\
 G & 12.823 $\pm$ 0.003\\
 A\textsubscript{g} & 0.16 $\pm$ 0.03 \\
 G\textsubscript{BP} & 13.066 $\pm$ 0.008\\
 G\textsubscript{RP} & 12.430 $\pm$ 0.006 \\
 (G\textsubscript{BP}-G\textsubscript{RP}) & 0.636 $\pm$ 0.001\\
 $T$\textsubscript{eff} & $6647^{+60}_{-54}$\\
 $\log$(g) & $3.83^{+0.04}_{-0.06}$ \\
Metallicity ([M/H])& $-0.51^{+0.90}_{-0.70}$ \\
 \hline
\end{tabular}
\end{table}
The \textit{Gaia} DR3 \citep{gaia2023} properties of ASASSN-24fw are summarized in Table (\ref{table:gaia}). The G-band absolute magnitude $M_{G}$ was determined from the extinction ($A_{g}$) corrected apparent G magnitude by subtracting the distance modulus, which was calculated from the distance of 1044$~\pm~$17 pc derived from the trigonometric parallax. Here, we have also included the 3D peculiar velocity computed using the equations by \cite{reid2009}, and \cite{poleski2013}. We find that ASASSN-24fw belongs to the Milky Way's thick disc distribution. We use {\it galpy} \citep{bovy2015}, a Python package to trace the orbit of ASASSN-24fw in Milky Way in the past 50 Ma. {\sc galpy} supports orbit integration in a variety of gravitational potentials, evaluating and sampling various distribution functions, and the calculation of action-angle coordinates for all static potentials. We find that the orbit of ASASSN-24fw does not intersect with any star formation region. Therefore this tells us that although ASASSN-24fw has a persistent infrared excess, it is unlikely to be a Young Stellar Object (YSO) and is more likely to be on main-sequence. 
\subsection{Search for intrinsic variability}
We searched for variability related to inherent stellar phenomena like star-spots, accretion, or rotation which is high in pre-main-sequence (PMS) stars in the TESS lightcurve of ASASSN-24fw. We find a peak at 0.41 d with a very small false-alarm probability. But the phase-folded lightcurve about this period is flat indicating that the frequency may not have any real significance. Since we do not have TESS lightcurve during dimming episode, we could not check the during dimming periodogram. But, instead we used ASAS-SN lightcurve 
to find repeating signals in the lightcurve of both the pre- and during dimming phase. Here we saw a peak at 1 day in both the periodograms 
which is actually common in ground-based photometry, and does not reflect any variability related to the star. 
As YSOs have high magnetic field, accretion and star spot activity, our analysis suggests that ASASSN-24fw is unlikely to be on pre-main-sequence and could be more evolved.
\subsection{Fundamental stellar parameters using Stellar Evolutionary Models applied to pre-dimming SED}
\label{ssec:star}
In the pre-dimming SED, we clearly see an infrared excess particularly in the mid-infrared \textit{WISE} data points (the first panel of Figure (\ref{fig:sed0})). 
We fit this SED with Castelli-Kurucz models\footnote{We have used VOSA \citep[VOSA v7.5,][]{bayo2008} \url{https://svo2.cab.inta-csic.es/theory/vosa/} to perform this analysis} \citep{castelli2003} and find the best fitting model has [T\textsubscript{eff},log(g),[Fe/H]] = [6250,5.0,-0.5]. This model is shown in cyan color in all the panels of Figure (\ref{fig:sed0}).

We then applied the Bayesian inference code of \citet{delburgo2016, delburgo2018} on a grid of stellar evolution models constructed from the PARSEC v1.2S tracks \citep{bressan2012,chen2014,chen2015,tang2014} to derive the fundamental stellar parameters of ASASSN-24fw. This choice is supported by the good statistical match obtained for detached eclipsing binaries, particularly for stars on the main-sequence, where the measured dynamical masses and corresponding predictions are consistent on average to within \,4\% \citep{delburgo2018}.

The grid of PARSEC v1.2S models arranged in this code consisted of ages ranging from 2 to 13,800 Ma in steps of 5\%, and [M/H] from -2.18 to 0.51 in steps of 0.02 dex. We adopted the photometric passband calibrations of \citet{riello2021}, and the zero points of the VEGAMAG system. We fed our code by the following three input parameters: the absolute magnitude $M_{G}$, the color $G_{\rm BP}-G_{\rm RP}$ (also adjusted for reddening and extinction using the values provided by Gaia Data Archive \citep{gaia2023}), and the iron to hydrogen abundance (assumed to be solar [Fe/H]=~$0.00~\pm~0.20$ dex).

Due to the presence of inherent infrared excess, as expected for YSOs, we also adopted a prior in the code that ASASSN-24fw is on PMS and recomputed the fundamental stellar parameters. In this case, we find that all the parameters except age of the star remain similar. We find that if the star is on the PMS, its age is 7 Ma and mass is 1.47$\pm$0.05 M\textsubscript{$\odot$}. Conversely, if it is on the main-sequence, its age is 2.75 Ga and mass is 1.69$\pm$0.08 M\textsubscript{$\odot$}. The other parameters are similar to each other in the uncertainty range. These results are listed in Tables Table (\ref{tab:stellar_parameters} and \ref{tab:stellar_parametersPMS}).

In Figure (\ref{fig:sed0}), we show the two different Kurucz models corresponding to the stellar atmospheric parameters that we just obtained. In addition to the cyan colored model that we discussed earlier, the {\it green} colored Kurucz model corresponds to [T\textsubscript{eff}, log(g), [Fe/H]] = [6000, 4.0, -0.5] and the {\it red} colored Kurucz model corresponds to [T\textsubscript{eff}, log(g), [Fe/H]] = [6500, 5.0, 0.0] i.e. the two models that we obtained by using the code of \cite{delburgo2016,delburgo2018}.

\subsection{Blackbody fit to the pre-dimming SED}
The emission in the mid-infrared and longer wavelengths comes from the circumstellar dust. This dust particles absorb the incident stellar flux and re-emit into the longer wavelength regions. Thus the nature of the excess flux can tell us about the nature of the circumstellar material. We can assume that the dust particles act as perfect black bodies and model the deviant points corresponding to the excess flux in the SED using a solid blackbody model. In Figure (\ref{fig:sed0}), it is visible that the excess flux actually manifests from the 2MASS $K_{s}$ band. So instead of a single component, we fit a two-component blackbody model. The warm component of our best fitted models gives us an effective temperature of $\simeq$ 1100K for the primary and $\simeq$ 400K for the secondary colder component. An effective temperature of 1100K, which is well below dust sublimation temperature, indicates a warm circumstellar material within 0.2 au radius from the star which is well below the dust sublimating temperature while an effective temperature of 387K hints at a colder disc located beyond 1.0 au from the star may consist of larger and rocky bodies. The two component black-body model is shown as a black dashed line passing through the WISE points.

\subsection{SED using the spectral observations during dimming}
In the second panel of Figure (\ref{fig:sed0}), we show the SED of the star during dimming
formed using our observed spectra. We see that all the three Kurucz
models fit well to the continuum but the main difference comes
while checking the fit to the peak of the SED especially Ca II H\&K
absorption lines which are deepest among all the lines. In the same plot, we have also shown an M8 brown
dwarf NIR spectrum template overplotted on top of the observed
Palomar NIR spectrum and the residual flux after subtracting the three model spectra from the observed spectra. This NIR template spectrum is taken from
the NIRSPEC Spectroscopic Brown Dwarf Survey \citep{mclean2007} which has a collection of spectral templates for the M, L, and
T type brown dwarfs. We can clearly see that the template matches
the initial part of the near-infrared continuum and there is an excess infrared flux from the end of H-
band. This hints towards the elevated levels of infrared excess during
dimming perhaps caused by circum-planetary rings to the brown
dwarf. We note that this excess is in the near-infrared wavelengths unlike the WISE wavelengths which span the mid-infrared region. Thus, the excess that we observe during dimming is most likely to be related to the occulter rather than the blackbodies that we fitted to the WISE points. Since we do not have any mid-infrared observations during
dimming, we cannot compare the pre-dimming and post-dimming
fluxes for mid-infrared wavelengths. 

\subsection{Spectral Analysis}\label{section:spectralanalysis}
The spectra that we have obtained were during the dimming episode of ASASSN-24fw. Thus they should be a mixture of the stellar spectra and the occulter spectra. It is therefore possible that, along with the continuum, the occulter may have affected the lines in the stellar spectra. For example, the variation in the hydrogen series lines and the Ca II H$\&$K lines in Figure (\ref{fig:lines}) in the LRIS spectrum taken on 31 October 2025 and HiRES spectrum taken on 28 January 2025 indicates that the occulter may be affecting the stellar spectra. Further, in the HiRES spectrum, we see that the age indicator Li line is not detected (or very weak at the level of noise), Fe, and Mg lines are moderately detected. These features as well as the shape and depth of the Ca II H$\&$K lines are normally expected for an F-type main-sequence star. Unlike YSOs, we do not see any emission line in our spectra. However, we see a central "reversal" feature or $w$ shape in the H$\alpha$ absorption line of the HiRES spectrum. Whether this is astrophysical or due to spectral noise is not very clear from the single spectrum. However, to find the strength of this line, we calculated the equivalent width of the absorption ($EW_{\rm absorb}$) and central emission ($EW_{\rm emission}$) component of the H$\alpha$ line. For this, we fitted a Lorentzian profile to the absorption line by masking the emission line corrected for barycentric motion. We find that $EW_{\rm absorb}$ $\sim$ 93{\AA} and $EW_{\rm emission}$ $\sim$ 7.8{\AA} and the radial velocity of the emission line in the heliocentric frame is -7.14 km/s. In contrast, \cite{zakamska2025} find that a two gaussian fit to their H alpha emission yields an EW $\simeq$ 1.98 {\AA} and the velocity centroid of the emission line is $\simeq$ -19 km/s in the heliocentric frame. This value is different from ours because the epoch of spectra as well as instruments are different.

\begin{table}[t]
    \centering
    \caption{Fundamental stellar parameters of ASASSN-24fw, inferred to be on the Main-Sequence.  \label{tab:stellar_parameters}}
    \begin{tabular}{lc}
    \hline\hline 
        Parameter & Value \\
    \hline
        $\Pi$ ($mas$) & 0.958 $\pm$ 0.015 \\   
        \hline
        $M_G$ (mag) & 2.57 $\pm$ 0.03\\       
        $G_{BP}-G_{RP}$ (mag) & 0.73 $\pm$ 0.04 \\  
        {[Fe/H]}    & 0.00 $\pm$ 0.20\\
        \hline
        Effective temperature (K) & 6146 $\pm$ 99 \\
        Radius ($R_{\odot}$) & 2.28 $\pm$ 0.08\\
        Mass ($M_{\odot}$) & 1.47 $\pm$ 0.05\\
        Mean density ($\rho_\odot$) & 0.125 $\pm$ 0.014\\
        Surface gravity ($\log g$, cgs) & 3.89  $\pm$ 0.03\\
        Luminosity ($L_{\odot}$) & 6.67 $\pm$ 0.22\\
        Bolometric magnitude (mag) & 2.68 $\pm$ 0.04\\
        Age (Ga) & 2.75 $\pm$ 0.24 \\
    \hline
    \end{tabular}
\end{table}

\begin{table}[t]
    \centering
    \caption{Fundamental stellar parameters of ASASSN-24fw, if assuming that it is on the PMS.}
    \label{tab:stellar_parametersPMS}
    \begin{tabular}{lc}
    \hline\hline 
        Parameter & Value \\
    \hline
        Effective temperature (K) & 6076 $\pm$ 121\\
        Radius ($R_{\odot}$) & 2.35 $\pm$ 0.11\\
        Mass ($M_{\odot}$) & 1.69 $\pm$ 0.08\\
        Mean density ($\rho_\odot$) & 0.132 $\pm$ 0.013\\
        Surface gravity ($\log g$, cgs) & 3.927  $\pm$ 0.022\\
        Luminosity ($L_{\odot}$) & 6.74 $\pm$ 0.23\\
        Bolometric magnitude (mag) & 2.67 $\pm$ 0.04\\
        Age (Ma) & 7.0 $\pm$ 1.8\\
    \hline
    \end{tabular}
\end{table}
\section{Nature of The Occulter}
\subsection{Lightcurve Modelling}\label{section:lightcurvemodel}
While we initially focused on characterizing the star, we now focus on understanding the occulter. The shape of the lightcurve can tell us a lot about its nature eg. small and regular dips can be attributed to exoplanets, while the irregular episodes of dimming can be attributed to patchy or fragmented circumstellar material. However, a long period and large scale dimming requires an object comparable to the size of the star. A single dark body comparable to the size of the star is physically impossible. But, the occulting body could be considered as a system of central body with massive circumplanetary rings and depending on the orientation of the ring system, a large portion of the star light can be blocked for longer duration. We proceed with lightcurve modelling in two steps; first we model the lightcurve using a parametric template that matches the shape of the lightcurve. This helps us to derive some basic parameters like the center of dimming, characteristic timescale, and lightcurve depth. In the next step, we can find a ring system geometry which can best explain the shape of the lightcurve. Here the information from the first model helps in setting the priors for Bayesian analysis used to maximize the likelihood of the fit. This is explained in the following sections. 
\subsubsection{Transit-Profile model}\label{section:lcmodel}
\begin{figure*}
\includegraphics[width=\textwidth]{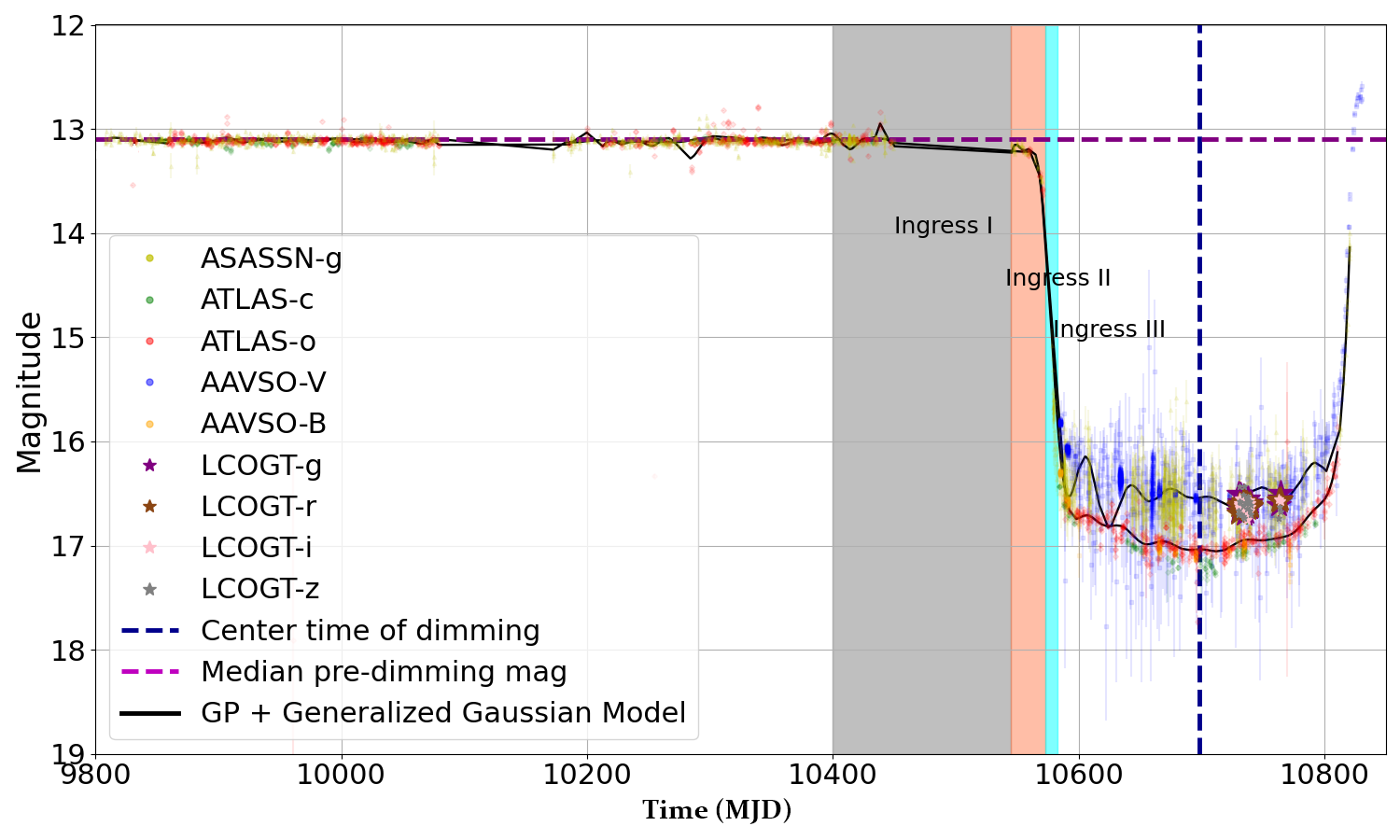}
\includegraphics[width=\textwidth]{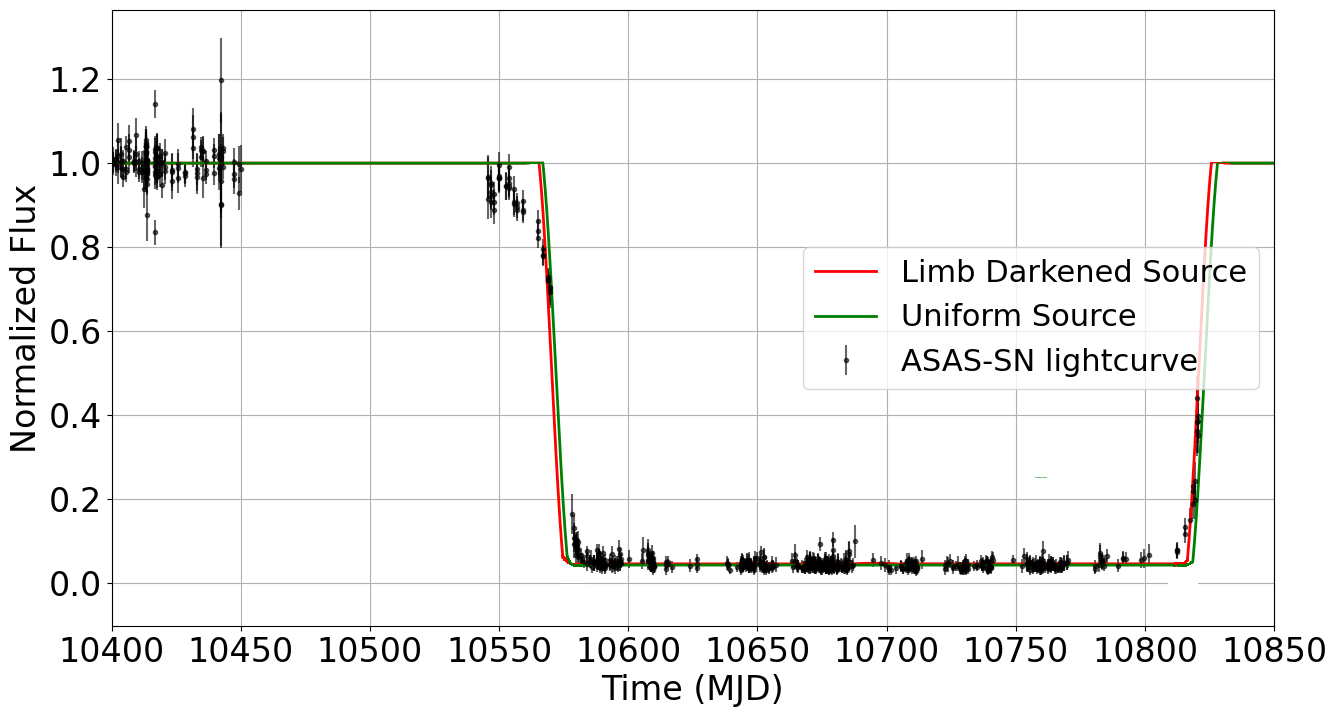}
\caption{\textit{Top}: The {\it Transit-Profile} model fitted to the lightcurve of ASASSN-24fw. This helps us to statistically estimate the maximum depth of the dimming, duration of the dimming and quantify the shape of the inverted hat shape of the lightcurve. Here we notice the different slopes in the lightcurve which indicate that the dimming started with a $\simeq$ 150 days gradual decrease in the brightness, followed two stages of ingress. The wiggles or the "micro-trends" in the model fitted to the ASAS-SN lightcurve are because of the Gaussian Process regression trying to find a mean model through the distributed data. They do not appear in the fit to the ATLAS lightcurve due to difference in the distribution of the noise and data points but the "macro-trends" are still highlighted. \textit{Bottom}: The occulting disc model generated using a central body with massive and optically thick circumplanetary rings, traversing the diameter of the star with a constant velocity and producing symmetrical lightcurve is shown. Although it misses the small features that the former model could show, from this model, we get an estimate on the geometry of the occulting system.}
\label{fig:ggdfit}
\end{figure*}

As the nearly symmetric lightcurve (see Figure (\ref{fig:ggdfit})) looks like an inverse hat, we fit the transit profile of the lightcurve using a parametric \textit{Generalised Gaussian}, defined by:
\begin{equation}
f\left(t\right) = 1 - A_{GGD} \exp\left(-\left| \frac{t - t_0}{\tau} \right|^\beta\right)
\label{eq:ggd}
\end{equation}
where $f$($t$) is the model brightness, $A_{GGD}$ is the maximum depth of the dimming, $t_{0}$ is the time of center of dimming, $\tau$ is the duration of the dimming, and $\beta$ is a parameter that defines the shape of the dimming. In case of ASASSN-24fw, $\beta$ should be high because of the flatness of the occulted lightcurve. ASASSN-24fw dropped to $\sim$ 16.75 magnitudes during dimming which is close to the detection limit of 17.50 mag of ASAS-SN camera. Therefore, the lightcurve would be affected by noisy data points. In order to model the long-term trends in the lightcurve and to handle the noisy data points, in the next step we used the non-parametric \textit{Gaussian Process} regression model. This was implemented through a Python package called {\it george} \citep{ambikasaran2015}, with a squared exponential kernel (\textit{ExpSquaredKernel}) given by:
\begin{equation}
k\left(t_{i},t_{j}\right) = A_{\rm GP}^{2}exp\Big(-\Big|\frac{t_{i}-t_{j}}{2l^{2}}\Big|\Big)
\label{eq:gp}
\end{equation}
where $A_{\rm GP}$ and $l$ are the kernel hyper-parameters. While the former hyper-parameter is the amplitude of the correlated noise in the whole lightcurve, the latter is the characteristic timescale of the noise in the whole lightcurve. Using these two hyper-parameters, {\it george} will try to find a mean model through the distributed data points. In this process, it can highlight any trends in the data which can otherwise be missed. Therefore, we combine these two models to capture the dimming and non-periodic trends in the lightcurve. We call our combined model as {\it Transit-Profile} model. We also performed the Bayesian analysis to maximize the likelihood of the fit and calculated the posterior distribution of our parameters. For this, we used the {\sc emcee} \citep{foreman2013} Python package which implements the {\it Affine Invariant Markov Chain Monte Carlo (MCMC) Ensemble sampler} algorithm introduced by \cite{goodman2010}. 

Our best-fitted {\it Transit-Profile} model is shown in the first panel of Figure (\ref{fig:ggdfit}) and the parameters with their 16\textsuperscript{th}, 50\textsuperscript{th}, and 84\textsuperscript{th} percentile values listed in Table (\ref{table:ggdparams}). This fit tells use few things; first, the star started dimming gradually $\simeq~$150 days (roughly in March 2024) much before the actual alert was issued during the \textit{ingress}. The second phase of dimming started around 02 September 2024 continued up to 13.66 magnitudes around 16 September 2024. Finally, the third phase of ingress caused a drastic and sudden drop in magnitude and the brightness reduced to $\sim$ 16.70 around 02 October 2024. This indicates that there are multiple layers of varying optical depth of material around the occulting body, possibly hinting towards a gas-giant body with optically thick rings. To check whether the gradual decrease in brightness is due to systematics, we downloaded the ASAS-SN lightcurves for all the stars within 3$\arcmin$ radius of ASASSN-24fw that had magnitudes 11 < $G$ < 15. We found four such stars. But they were invariant at 0.10 mag level (3$\sigma$ level). This rules out the fact that the Phase I of the ingress is due to telescope systematics but is due to the outer layers of the occulter system.

To check if our {\it Transit-Profile} model is really capturing various trends in the data or is just over-fitting the lightcurve. For this, we repeat the transit-profile model fitting on the ATLAS survey lightcurve. The photometric limit of ATLAS camera is deeper than the ASAS-SN camera and hence the data would be less noisy. We see that the wiggles or the "micro-trends" in the fit have reduced but the three phases or the "macro-trends" of ingress are similarly fitted (see the red and green colored data points and the model passing through them in the Figure (\ref{fig:ggdfit})). This indicates that the long-term or the "macro-trends" in the lightcurve are really present and discovered using our method.

\begin{table}
\centering
\caption{The parameters of the {\it Transit-Profile} model.}
\label{table:ggdparams}
\begin{tabular}{ |c|c| } 
 \hline
 Parameter & Value  \\ 
 \hline\hline
 $A_{\rm GGD}$ & 0.89\\
 $A_{\rm GP}$ & 137.73 \\
 $\tau$ (days) & 272.35 \\
 $t_{0}$ (days) & 2460697.94\\
 $\beta$ & 4.72 \\
 $l$ & 683.94 \\
 \hline
\end{tabular}
\end{table}
\begin{table}
    \centering
        \caption{Occulting disc model parameters with 16\textsuperscript{th}, 50\textsuperscript{th},, and 84\textsuperscript{th} percentile values.}
    \label{tab:pyppluss}
    \begin{tabular}{|l|c}
        \hline
        Parameter & Value \\
        \hline\hline
        $R_{\rm in}$ ($R_{\star}$) & $0.11^{+0.11}_{-0.08}$ \\
        $R_{\rm out}$ ($R_{\star}$) & $16.71^{+7.12}_{-2.81}$ \\
        $i$ (degrees) & $39.85^{+8.49}_{-13.19}$ \\
        $\phi$ (degrees) & $49.03^{+25.73}_{-25.73}$ \\
        $v^{'}_{T}$ ($R_{\star}$)day\textsuperscript{-1}& $1.01^{+0.08}_{-0.09}$\\
        $\tau_{R}$ & $0.96^{+0.01}_{-0.01}$\\
        $t_{0}$ (HJD-2450000 days) & $10698.50^{+17.62}_{-2.06}$ \\
        \hline
    \end{tabular}
    \label{table:pyppluss}
\end{table}

\subsubsection{Occulting disc model}\label{section:discmodel}
We now understand that the dimming is not cause by a single solid body but by a system with large area, varying density and  optical depth. So in the next step, we try to model the lightcurve by a point like physical body having a massive circumplanetary disc similar to that of $\epsilon$ Aur type systems (for example, \citep{rodriguez2015, rodriguez2016, rappaport2019, vanderkamp2022}. We call this model the {\it Occulting disc} model. To compute this model, we use the Python package named {\sc pyPplusS}\footnote{\url{https://github.com/EdanRein/pyPplusS}} \citep{rein2023}. In our model, the dimming is caused by elliptical projection of the massive circumplanetary rings on to the star. Unlike the short dips caused by transiting planets which cover a small portion of the stellar surface, the large scale discs block a significant portion of the stellar area and hence require a different approach to calculate the stellar brightness during dimming. {\sc pyPplusS}'s approach is to use the {\it Polygon-Segments} algorithm which efficient and stable algorithm.

Typically, the parameters for computing a model in {\sc pyPplusS} are: planet radius ($R_{\rm p}$), the inner diameter of the ring ($R_{\rm in}$), the outer diameter of the ring ($R_{\rm out}$), the inclination angle of the ring from the orbital plane of the exoplanet ($i$), the azimuth angle by which the ring is rotated about the z-axis (or the exoplanet's spinning axis) ($\phi$), opacity ($\tau_{R}$), and the limb-darkening coefficients of the star. We also consider a uniform source and do not find any significant difference in the generated model. In addition, the code requires time dependent coordinates of the exoplanet's center $x_{p}$ and $y_{p}$. We modified our modelling approach by considering a point like exoplanet surrounded by a massive ring system with sharp edges which is comparable to or larger than the stellar size. We further simplify our model by considering $x_{p}$ = $v_{t}\times$($t-t_{0}$), and $y_{p}$ = impact parameter $b\times R_{\star}$ (or b $\sim$ 0). Here $v(t)$ is the transverse velocity of the system which is assumed constant throughout the event, and $t_{0}$ is the time of the closest approach. Our simplification is made in a way that the exoplanet is traversing the star in an edge-on orbit and across its diameter. Thus our model parameters are: $R_{in}$, $R_{out}$, $i$, $\phi$, $\tau_{R}$, $v_{t}$ and $t_{0}$.

We explored this parameter space by looking for models that could produce the $\simeq$ 275 day long transit event, while also minimizing the $\chi^{2}$ values. Here also we have made use of the Python package {\sc emcee}. Obviously, the models that would produce such long transits need to have low transverse velocities that likely indicate the large semi-major axis between the occulter and the primary star. Our best-fitted model is shown in the lower panel of Figure (\ref{fig:ggdfit}) (uniform source in green color and limb-darkened source in red color) and the fitted parameters with their 16\textsuperscript{th}, 50\textsuperscript{th}, and 84\textsuperscript{th} percentile values are shown in Table (\ref{table:pyppluss}). We see that our model fits the lightcurve well, but does not accurately describe the lightcurve because of a couple of things. First, the assumption that we made that rings have sharp edges is not entirely correct, and second, the trajectory may not be traversing through the diameter of the star. Nevertheless, our model gives an estimate on the geometry of the system. We see that our planet is surrounded by optically thick rings with the outer diameter being roughly 16~$R_{\star}$ or $\sim$ 0.17 au. This system is inclined to the orbital plane by nearly $40^\circ$, rotated about the rotational axis by nearly $50^\circ$
 and v(t) is about 1~$R_{\star}$ per day or $\simeq$ 17 km/s. We note that in practice {\it v(t)} is expected be degenerate with other model parameters. However, in our simplistic approach, we have constrained our model by fixing the size of the exoplanet to a point and assuming an edge-on trajectory traversing the diameter. This constrains the parameter space to be explored and resolves the degeneracies. Moreover, the orbit of the planet may be inclined at an angle. But the projection of the rings on the star blocks an area of the star equal to a portion of an ellipse. This area is proportional to the amount and duration of the light getting blocked. Since this is a massive and long duration dimming, the area of the star blocked is much larger than the ones blocked by a single transiting exoplanet. So we are not modelling the transit by an exoplanet but the transit by projected rings that are blocking the stellar light.

As a check on our computed v(t) using lightcurve model, we try another method to compute the transverse velocity of the occulter. For example, \cite{van2014} calculate the transverse velocity based on the maximum gradient of the lightcurve. It requires the stellar linear limb-darkening parameter ($u$), of the star which can be computed using the {\sc jktld} code of \cite{southworth2015}. This code uses the stellar model parameters of the star to linearly interpolate the tables from \cite{sing2010} and calculates {\it u} for the required band passes. For ASASSN-24fw, we obtain {\it u} = 0.48 for the g'-band data of ASAS-SN survey. Using this value of $u$, the steepest time gradient of the light curve, $\dot{L}$, and taking R=$R_{\star}$ (from Table (\ref{tab:stellar_parameters})), we can get another estimate on $v_{t}$:
\begin{equation}
v_{t} = \dot{L} \times 2.30 R_{\odot} \times \pi \left( \frac{2u - 6}{12 - 12u + 3\pi u} \right)
\label{eq:vt}
\end{equation}

From the lightcurve, we get the maximum value of the gradient $\dot{L}$ = 4.97$\pm$0.43 $\times$10\textsuperscript{-6}~L\textsubscript{$\star$}day\textsuperscript{-1}, (where L\textsubscript{$\star$} is normalized to L\textsubscript{$\odot$}). Therefore, we obtain $v_{t}$ $\simeq$ 13.74 $\pm$ 2.49 km/s. This is close to the orbital speed of $\simeq$ 11.5 km/s derived from Kepler's third law, considering an orbital period of $\simeq$ 16000 days \citep{nair2024} and an orbital semi-major axis of $\simeq$ 17 au. This is also not very different from the velocity that we get from our lightcurve model. Then the occulter radius ($r_{disc}$) = $v_{t}$$\delta{t}$/2 $\simeq$ 0.16 $\pm$ 0.02 au.
This is similar to the outer radius of the disc that we get from our model $\simeq$ 0.17 au. Since for a stable disc, its radius should be smaller than the Hill radius of the secondary, the lower limit we get on the mass of the secondary is $M_{2}$ $\simeq$ 3$M_{\star}$($r_{\rm disc}/a$)\textsuperscript{3}> 3.42 $M_{J}$. Therefore, we can safely assume that the secondary is most compatible with a super-Jupiter or a brown dwarf with optically thick rings.

\subsection{Extinction calculations using LCOGT data}\label{section:extinction-calc}
Our aim of obtaining multi-band LCOGT observations was to monitor the variation in color of ASASSN-24fw during dimming. The variation in color can then be used to compute the specific reddening vector and hence the composition of the occulting material. 
Unfortunately, our post-dimming baseline of ASASSN-24fw using LCOGT observations in the four SDSS filters is not very long. However, it is still enough to compute the reddening vector and hence constrain the primary composition of the occulting material. We measure mean colors for ASASSN-24fw of $g^{\prime}$-$r^{\prime}$ = 0.40~$\pm$~0.07~mag, $r^{\prime}$-$i^{\prime}$ = 0.16~$\pm$~0.05~mag, and $r^{\prime}$-$z^{\prime}$ = 0.28~$\pm$~0.10~mag. For comparison, we estimate the ISM extinction using the Python package {\sc extinction} \citep{barbary2021} to extinguish the stellar photosphere model using the Fitzpatrick '99 \citep{fitzpatrick1999} extinction law and obtain synthetic photometry in the necessary filter bands using the Python package {\sc pyphot} \citep{fouesneau2025}. 
We find that the photospheric colors for ASASSN-24fw should thus be $g^{\prime}$-$r^{\prime}$ = 0.42~mag, $r^{\prime}$-$i^{\prime}$ = 0.25~mag, and $r^{\prime}$-$z^{\prime}$ = 0.17~mag, assuming a distance of 1044~pc. The observed reddening is therefore not consistent with ISM extinction, being significantly stronger at longer wavelengths than expected.

We generate color-color plots using $g^{\prime}$, $r^{\prime}$, $i^{\prime}$ and $z^{\prime}_{s}$ bands and use {\sc optool} \citep{dominik2021} to calculate the opacities $Q_{\rm ext}$ (= $Q_{\rm abs}$ +$Q_{\rm sca}$) values for a range of dust grain compositions and two size distributions. We used the built-in refractive indices from {\sc optool} to generate the $Q_{\rm ext}$ values for materials including astronomical silicate \citep{draine2003}, various crystalline and amorphous iron-magnesium silicates \citep{dorschner1995, jaeger1998, fabian2001, suto2006}, carbon-bearing species \citep{draine1993, henning1996}, quartz \citep{kitamura2007}, and iron particles \citep{henning1996}.

We consider a log-normal grain distribution with a width of 0.5 dex and a free peak size, considering contributions from sizes between 0.1 and 1.1$\mu$m in steps of 0.01$\mu$m when calculating $Q_{\rm ext}$, determined from the complex refractive indices assuming Mie theory \citep{mie1908}. For a Sun-like star, the typical blowout size for a dust grain is $\simeq$ 0.5$\mu$m \citep{krivov2010}, so we include transient dust with this range of grain sizes. The optical depth $\tau$ and extinction of a dust cloud that fully covers the stellar disc are then calculated from the $Q_{\rm ext}$ values for each combination of composition and size distribution, which are then compared to the measured reddening vectors by least-squares fit to identify those which best match the observations. For ASASSN-24fw, we find that no combination of compositions available in {\sc optool} and grain size produces a reddening vector that fits \textbf{all} the data. We can match individual sizes and compositions to separately fit the bulk of measurements in the $g^{\prime}$-$r^{\prime}$ vs $r^{\prime}$-$i^{\prime}$ and $g^{\prime}$-$r^{\prime}$ vs $r^{\prime}$-$z^{\prime}$ color-color plots but no single material adequately reproduces the reddening vectors in both. \citet{toribio2025} found silicate materials around 1~$\mu$m in size could reproduce the $g^{\prime}$-$z^{\prime}$ opacity gradient; we find astronomical silicate dust grains around 0.4 to 0.6~$\mu$m in size is the best, albeit imperfect, overall match to the reddening.

\section{Discussion}
\subsection{Is ASASSN-24fw an R Corona Borealis (R CrB) type variable?}\label{section:rcb}
R CrB stars also show semi-regular pulsations with periods of multiple tens of days and amplitudes up to a few tenths of a magnitude \citep{percy2023}; in contrast, as mentioned before, the star does not show significant variability in its TESS light curves. R CrB type variables come under the parent class of hydrogen-deficient stars, are found in a temperature range similar to that of ASASSN-24fw, and are known to show similar long (although sporadic) dimming behaviour \citep{clayton1996, clayton2012, schaefer2024, crawford2025}. R CrB stars are also known to exhibit mid-IR excess due to warm circumstellar dust shells \citep{chesneau2014, rao2015, tisserand2020}. The spectrum of R CrB stars are known for their hydrogen-less erratic and asymmetric dimming profiles and are dominated by mostly helium, nitrogen, and carbon enrichment in their atmospheres \citep[see ][]{asplund2000, pandey2021}. We found deep hydrogen absorption lines and no helium emission or P Cygni profile in our Keck HiRES optical spectrum (the second plot in Figure (\ref{fig:sed0})) and also the post-dimming Magellan MIKE Spectrum (See Figure (\ref{appendixfig:a})). Moreover, no carbon absorption features or other spectroscopic R CrB signatures are seen in our optical spectrum \citep[see ][]{pandey2021}. In this study, we have not performed a detailed spectral analysis in search for other elements characteristic of hydrogen deficient stars (eg. Germanium \citealt{saini2025}). However, this will be performed later when once we obtain a post-egress spectrum. Nevertheless it is highly unlikely that ASASSN-24fw is a R CrB type of variable. 

\subsection{Is ASASSN-24fw a YSO?}\label{section:ttauri}
YSOs are of F- or G- spectral type, usually found near molecular clouds, and often show optical variability and accretion signatures due to high magnetic fields and active circumstellar zone comprised of an protoplanetary/proto-stellar material around them \citep{joy1945, gahm2008, zsidi2025}. All YSOs have infrared excess and the ones that show optical variability are called as T Tauri type of variables (or "dippers"). As discussed earlier, this variability is manifested as irregular and non-periodic dips and is linked with clumpy circumstellar material \citep{cody2014, ansdell2016, ansdell2019}. It has been found by \cite{ansdell2016}, and \cite{nunez2017}, that a positive correlation exists between the depth of dips and the amount of infrared excess.

The circumstellar material around a T Tauri type of variable star is divided in to two components: (1) the inner component, which is warmer and contains finer dust that accretes on to the host and (2) the outer component which is mixture of gaseous and solid material. Together they may manifest as two solid bodies causing infrared excess in the SED similar to what we see here (e.g. \citealt{varga2018} and references within). In ASASSN-24fw we see an infrared excess as well as a deep but long dimming causing a flat base unlike the irregular shaped lightcurve. Therefore the circumstellar material causing infrared excess in ASASSN-24fw appears to have spread out rather than concentrated in clumps. This is supported by the fact that the mid-infrared NEOWISE lightcurve is also non-variable.

Although the shape of the SED and the mass of ASASSN-24fw is similar to that of a T Tauri star, in the only pre-dimming low-resolution spectrum that we have from \textit{Gaia} DR3, no H$\alpha$ emission line appears which is an indicator of accretion activity. Similarly, in the low resolution during-dimming spectrum obtained from Keck, we do not see any emission signatures. In our Keck high-resolution spectrum taken in January 2025, we see several deep hydrogen absorption lines. If the Ca II H$\&$K lines are unaffected by the occulter, then their strength resembles to that of an F-type main-sequence star. However, in the same spectrum, we see that the H$\alpha$ absorption line is shallower unlike other hydrogen series lines and has a $w$ shape resembling a central emission feature (as already discussed).

If the $w$ shape is not due to spectral noise then a component of the occulting material is Hydrogen rich and re-emitting the incident stellar radiation into H$\alpha$ region. The otherwise flat lightcurve of ASASSN-24fw only shows deep occultation every 43.8 years \citep{nair2024} which indicates that the occulter is large and optically thick. This is unlike any other lightcurve of a T Tauri star where we expect regular episodes of irregular dips.

In most cases the accreting circumstellar discs around YSOs are bright in X-ray and radio wavelengths owing to the strong magnetic fields and high temperatures from ongoing accretion activity \citep{feigelson1999, espaillat2019}. If the companion is a stellar remnant, we can also see X-ray emission from the system. We searched through known X-ray catalogs like {\it ROSAT 2RXS} \citep{boller2016}, {\it XMMDR10} \citep{webb2022}, {\it XMMSL2} \citep{xmm2018}, {\it Chandra CSC2} \citep{evans2024}, and {\it swift2SXPS} \citep{kosiba2024} and found no pointing at the location of ASASSN-24fw. Thus we do not have evidence that the ASASSN-24fw has a compact companion or there are X-rays emitted from accretion acitivity.
 
In Appendix (\ref{appendix:a}), we show a high-resolution post-dimming spectrum taken on 13 October 2025. Here we can see a deep H$\alpha$ line in absorption and not in emission. If ASASSN-24fw were an YSO, and the mid-infrared excess causing material was a proto-planetary material, we would expect strong H$\alpha$ emission line. It may be possible that accretion may be happening in a different location on the stellar surface and not along our line of sight. However, we do not see any reason as to why this accretion spot has to change post-dimming (as we have already rules out fast stellar rotation). This, and the absence of Li line, the age indicator for main sequence stars \citep{andres2021}, supports the fact that the infrared excess causing material is not proto-planetary and is likely to be a debris disc. Also as mentioned earlier, the peculiar velocity of ASASSN-24fw indicates that it does not appear to have interacted with any star forming regions at least in past 50 Ma and that it belongs to the older population of stars in thick disc. These things support the fact that ASASSN-24fw is unlikely to be an YSO.

\subsection{Nature and age of the system}
The larger distance, longer orbital period, large depth of dimming and longer eclipse duration of ASASSN-24fw makes it a unique, and observationally challenging system. The crude stellar analysis based on our available data suggests that the star is unlikely to be a YSO although it has a persistent infrared excess. It is likely that the star has an active planetary environment and the catastrophic collisions between planetesimals has caused a debris disc around it. Although this can be verified, the distance of $\sim$ 1kpc would make it challenging to resolve the discs. Nevertheless, mid-infrared studies from JWST can tell us about the composition of the circumstellar environment. Follow up series of observations from NIRCam or VLT SPHERE can infact be useful to measure the proper motion and confirm the companionship of the faint source that we found in our LCOGT images.

The lower limit on the mass of the secondary that we derived indicates that the secondary is either a super-Jupiter or a brown dwarf. There is no clear distinction between the lower-limit of brown dwarf mass and upper limit of super-Jupiter objects, as they are mainly categorized based on the formation mechanisms. For example, the  core-accretion theory \citep{pollack1996} can explain the formation of gas giant companions at separations $<$100 au, but gravitational instability or disc fragmentation \citep{boss1997} is required for heavier gas-giants. Although the number of planets formed by core accretion is more than those formed by gravitational instability, the increase in the detection of wide and massive planets as in case of ASASSN-24fw, suggests that both core-accretion and gravitational instability together could also be responsible for planet formation. This in fact requires further exploration and detecting and analyzing more such events would be a nice step in this direction.

One of the distinguishing factors between the super-Jupiters and brown dwarfs can be their discs. If the secondary is a super-Jupiter (with 5–13 M$_{Jup}$), it could host a large disc, provided it is young or resides in a debris-rich environment. Such a system could persist, especially at $\simeq$20 au where dynamical clearing is slow (for example, \cite{liu2023}). However, forming a super-Jupiter hosting a {\it optically thick} disc with a size of $\simeq$272 times the size of Jupiter is complicated as it can become gravitationally unstable owing to reduced lifetime. Instead, the companion being a more massive brown dwarf is more likely scenario.

Brown dwarfs are believed to form like stars via a gravitational collapse, and thus can retain the surrounding material \citep{stamatellos2009, stamatellos2015}, especially when the disc radius is well outside the Roche limit of the star. 
A disc of $\simeq$0.20 au radius is consistent with observed discs around young brown dwarfs (e.g., in star-forming regions like Taurus \citep{kutovic2021}). Such discs comprise of gas and fine dust (micron-sized grains) that scatter or absorb the stellar light effectively.
This material is possibly replenished, from the ongoing accretion \citep{stamatellos2015, zhu2015}.

Although replenishing process may be responsible for the observed disc, this process is efficient only in the first few million years of the formation. As the system evolves, various "outside" forces come affect the process. Thus, if we consider that the star is an older main-sequence star, then an age of about 2 Ga introduces new challenges for the disc survival. On such longer timescales, the brown dwarf’s disc is expected to be depleted due to dynamical and radiative processes. Moreover, as a mature disc looses its opacity due to dust coagulation or dispersal, a significant luminosity drop in ASASSN-24fw due to a older disc seems improbable. Although at larger separations, the brown dwarf’s disc would experience minimal stellar radiation, but the internal processes (e.g., viscous evolution) and external perturbations (e.g., passing stars) would likely deplete it long before 2 Ga.
Therefore, a massive, opaque circumplanetary disc that is surviving for so long on its own is unlikely unless it is replenished by some process (e.g. by accretion from the circumstellar environment which could be rich in material due to active collisions).

Another process for the replenishment is the Bondi–Hoyle–Lyttleton accretion \citep{hoyle1939, bondi1944} which would have played a role if the brown dwarf would have passed through an unusually dense region of the ISM (e.g., a molecular cloud or diffuse gas) in the past.

\subsection{Comparison with recent studies on ASASSN-24fw}
After the event ended in mid-2025, it validated the predicted egress by \cite{nair2024}. Soon thereafter, two groups \citet{toribio2025} and \citet{zakamska2025} reported their observations and conclusions. While both the studies agree on the large grain sized dusty component involved in the dimming event, and that ASASSN-24fw is an F-type star and evolved on a main-sequence, they differ on the interpretations of the event. But, \citet{toribio2025} also find degeneracy in the age (and hence the mass of the star) like us. They favour a model with the presence of a 0.25~$M\textsubscript{$\odot$}$ M-dwarf secondary and that a circumbinary disc instead of a circumplanetary disc responsible for the dimming. Further, as they do not find any color change during eclipse (using their multi-band LCOGT and UKIRT observations), and find time-varying polarization during dimming, they suggest that the occulting material is comprised of large ($>$20$\mu$m) carbonaceous or water ice grains rather than small silicate and CO\textsubscript{2} grains possible in circumbinary discs. 

\cite{zakamska2025} undertook spectral study of the dimming and suggest that ASASSN-24fw is an evolved main sequence star and the dimming was caused by a tidally truncated circumplanetary disc comprising of gas and dust of large sized grains. They also rule out shorter orbital periods using the DASCH survey and modern survey lightcurves. They provide mass of the occulter to be in between 19-76 ~$M_{\rm J}$ and the radius of the disc to be 0.7 au. Moreover, their analysis using deep NaI D absorption lines which are in excess than that expected from stellar photosphere alone suggest the possibility of circumsecondary winds. Unlike their spectra neither we nor \cite{toribio2025} found any metallic and H$\alpha$ emission\footnote{Though we see a central emission feature, it is below the continuum level. So we refer to the shape of H-alpha line as {\it w} shaped feature.} lines in our optical spectra. Interestingly, the optical spectrum of \cite{toribio2025} taken on 9 October 2024 i.e. just after the star became dim, shows deep H$\alpha$ absorption line, and our Keck LRIS spectrum taken on 29 October 2024 shows no H$\alpha$ line. But our Keck HiRES spectrum taken on 25 January 2025 shows a H$\alpha$ absorption line with a {\it w} shape or a central emission feature. Interestingly, \cite{zakamska2025} find H$\alpha$ emission line in their Magellan and GHOST spectra taken on 3 February 2025 and 5 March 2025 respectively, after our Keck HiRES observations. As the only available pre-dimming spectra is from Gaia DR3 which also shows H$\alpha$ absorption line, and that we have ruled out that ASASSN-24fw is a YSO, indicates that the emission feature is unlikely to be coming from the star and maybe associated with the hydrogen gas component in the occulting material which is re-emitting the absorbed stellar radiation into H$\alpha$ region. The value of $EW_{\rm absorb}$ that we find in our HiRES spectrum is $\sim$ 8\AA. This is different if compared with the values by \cite{zakamska2025} and therefore indicates a variable nature of the H$\alpha$ line.

\subsection{Future plans}
Long dimming events like ASASSN-24fw are open to interpretation just based on lightcurve and SED analysis. 
To improve our analysis, we plan to conduct post-dimming observations and determine the stellar spinning speed from the rotational broadening, metallicity and measure the Ca II H\&K and Li elemental abundances to estimate the age and evolutionary status of the star. 

We also plan to compare the out-of-eclipse spectra (shown in Figure (\ref{appendixfig:a})) with the during-dimming Keck HiRES and get a transmission spectrum to further characterize the composition of the occulter. However, at this stage, it is evident that ASASSN-24fw is an interesting object and that its nature is degenerate with its age and the occulter is massive in size and has rings made materials in varying densities. The post-dimming spectra already tells us that ASASSN24fw is not an YSO as the H$\alpha$ line is in absorption. It also tells us that Ca H\&K and Na Db lines are one of the deepest lines and the age indicator Li line is very weak. We would present our in-depth analysis on this in our upcoming papers.

We also plan to conduct out-of-dimming polarimetric study of ASASSN-24fw and nearby ISM to understand the amount of polarization and the nature of the circumstellar environment causing the infrared excess. Similarly, the sub-millimeter observations will be helpful in further constraining the circumstellar environment of ASASSN-24fw. The JWST observations using MIRI will help us to detect whether the composition of the circumstellar environment is dominated by the silicate or the carbonaceous dust. Infact if the survival of the massive circumplanetary disc is indeed due to Bondi–Hoyle–Lyttleton accretion, we could also see signatures of ISM molecules in the NIRSpec and MIRI spectrum. Finally, observations using high-angular resolution instruments on large telescopes would help us to check the companionship of the faint source detected in our LCOGT images. All the observations, would be helpful to create a complete picture about ASASSN-24fw and give us insight about the evolution of such systems, existence of gas-giants with optically thick rings and the environments in which they form. Systems like ASASSN-24fw would be of great importance for the upcoming extremely large telescopes.

\section{Conclusions}
We have analyzed the recent dimming event around the star ASASSN-24fw. It previously exhibited stable brightness over decades of observation and then abruptly started noticeable dimming on 02 September 2024. The star had a persistent mid-infrared excess pointing towards the presence of a circumstellar material. The estimated fractional luminosity of this material is $\simeq$ 10\%. This excess did not exhibit any periodic variation in the {NEOWISE} lightcurve prior to dimming suggesting that the material is spread out around the star and is not located in clumps. The blackbody fit requires two components with temperatures of 1100~K and 400~K. From our lightcurve modelling, we suggest that this material is unlikely to be responsible for the recent dimming event. 

We modeled the lightcurve in two stages: (1) using a combination of \textit{Generalised Gaussian Distribution} and \textit{Gaussian Process Regression} called the {\it Transit-Profile} model and (2) assuming a point like planet with massive and inclined rings traversing the diameter of the star called the {\it Occulting-disc} model. Our first model suggests that the dimming actually started in March 2024 much before the actual alert was issued, and it progressed in stages. This indicates that the outer disc came in contact earlier than the main body much before the actual ingress. The different slopes of the model indicate that the density of the occulting material is varying and inversely proportional to the radial distance from the central body. While our model overfits the "micro-trends" due to the noise in the ASAS-SN lightcurve during the dimming, fitting a similar model to the ATLAS lightcurve suggests that the macro trends in the lightcurve representing various stages of dimming are significant. Our second model is a very simplified assumption of the occulter being a massive and optically thick ring system around a gas-giant planet and passing across the stellar diameter. This model suggests that the inner diameter of the disc is $\simeq$ 0.11 $R_{\star}$ and the outer diameter of the disc is $\simeq$ 16 $R_{\star}$ or $\simeq$ 0.17 au (within the hill sphere). Together this system weighs at least 3.42 $M_{J}$ and is having a constant transverse velocity of roughly 17 km/s. It also tells us that the ring system is has an opacity of $\simeq$ 0.96 and is inclined to the orbital plane at an angle of 40\textsuperscript{o} and azimuthally rotated about the orbital spin axis at an angle of 50\textsuperscript{o}. This orientation casts a massive elliptical shadow on the star that is responsible for the dimming event. This model does not fit the ingress very well because our assumptions that the planet is traversing the diameter of the star and the transverse velocity may actually be different for different layers or disc material. Further, our {\tt OpTool} analysis using the post-dimming LCOGT data that we obtained from February to April 2025 indicates that the disc material mainly consists of astronomical silicate dust grains around
0.4 to 0.6 $\mu$m in size.

We modeled the pre-dimming SED of the star using \textit{VOSA} \citep{bayo2008} and also the Bayesian inference code developed by \cite{delburgo2016, delburgo2018}. Using the mean value of reddening towards the star, \textit{VOSA} gives us an age of roughly 8 Ma, mass around 1.6 M\textsubscript{$\odot$} and [T\textsubscript{eff}, log(g), [Fe/H]] = [6250, 5.0, -0.5]. In the latter code we consider two possibilities that the star is in the pre-main sequence phase, or on the main-sequence. Whilst most of the stellar parameters derived from these models are similar to what we get from VOSA, we get two different ages of 7 Ma and 2 Ga. Since this analysis is based on the pre-dimming photometry, an accurate age and composition of the star will also be determined from the post-dimming high resolution spectrum that we have obtained. We will also compare the during-dimming and post-dimming spectra to generate a transmission spectrum and measure and understand the composition of the occulter. While two models suggest that the star is young, we find that its kinematics belong to the thick disc distribution. It is not close to any star-forming region nor it has arrived from one in the past 50 Ma. We also check the TESS lightcurve of ASASSN-24fw. If ASASSN-24fw is a YSO or a young star, then its high magnetic field or star spot activity should be reflected in the TESS lightcurve. But the lack of any variability indicates that ASASSN-24fw is a older main-sequence star.

The post-dimming spectra from Keck LRIS and Keck HiRES resemble an F-type main-sequence star. 

This line was in absorption before January 2025 and in emission later. Our spectrum taken in late January 2025 shows a central-emission {\it w} shaped reversal feature which suggests that the H$\alpha$ line reversed from absorption to emission post January 2025. Further, in the post-dimming spectrum taken in October 2025, we do not find the H$\alpha$ line in absorption. Therefore, it is most likely that the emission feature is associated with the occulter and not with the accreting circumstellar environment of the ASASSN-24fw.

Based on the massive ring size that we get, the companion is more likely to be a brown dwarf rather than a gas giant planet. As such massive circumplanetary discs should not survive beyond $\simeq$ 100 Ma because the material would dissipate over time due to accretion, photo-evaporation, or dynamical clearing, the existence of such a massive circumplanetary disc that is causing periodic dimming every 43.8 years means an extraordinary mechanism like a recent collisional replenishment, a anomalous disc stability, or Bondi–Hoyle–Lyttleton accretion processes are playing/might have playedrole. This challenges our understanding of circumstellar disc lifetimes and brown dwarf environments and makes ASASSN-24fw interesting system for future studies. If the age is consistent with the 2 Ga interpretation then the long-term existence of the circumplanetary scenario will be confirmed, and would require new theories about disc survival, debris disc dynamics, or unseen dynamical influences in mature systems.

Events like ASASSN-24fw are unique, open to interpretation. Detailed analysis of such events are very important to understand stellar, circumstellar disc, and planetary evolution.

\section*{Acknowledgements}
We thank the referee(s) for their suggestions and feedback to improve this manuscript. JPM acknowledges research support by the National Science and Technology Council of Taiwan under grant NSTC 112-2112-M-001-032-MY3. CdB acknowledges support from the Agencia Estatal de Investigación del Ministerio de Ciencia, Innovación y Universidades (MCIU/AEI) under grant WEAVE: EXPLORING THE COSMIC ORIGINAL SYMPHONY, FROM STARS TO GALAXY CLUSTERS and the European Regional Development Fund (ERDF) with reference PID2023-153342NB-I00/10.13039/501100011033, as well as from a Beatriz Galindo Senior Fellowship (BG22/00166) from the MICIU. The Universidad de La Laguna (ULL) and the Consejería de Economía, Conocimiento y Empleo of the Gobierno de Canarias are also gratefully acknowledged for the support provided to CdB (2024/347). GH acknowledges support from the European Research Council (ERC) under the European Union's Horizon 2020 research and innovation program (grant agreement no. 695099). SZ acknowledges support from the Research Fellowship Program of the European Space Agency (ESA). SGD acknowledges generous support from the Ajax Foundation. FK acknowledges support from the Spanish Ministry of Science, Innovation and Universities, under grant number PID2023-149918NB-I00. This work was also partly supported by the Spanish program Unidad de Excelencia María de Maeztu CEX2020-001058-M, financed by MCIN/AEI/10.13039/501100011033. This publication makes use of VOSA, developed under the Spanish Virtual Observatory (https://svo.cab.inta-csic.es) project funded by MCIN/AEI/10.13039/501100011033/ through grant PID2020-112949GB-I00. VOSA has been partially updated by using funding from the European Union's Horizon 2020 Research and Innovation Programme, under Grant Agreement No. 776403 (EXOPLANETS-A). During the post-dimming spectroscopic observations, ASASSSN-24fw was observed as a backup target during the CNTAC program
CN2025B-33 (PI: W. Gieren, observer: B. Pilecki).

\textit{Software}: This work made use of {\it Astropy}\footnote{http://www.astropy.org} a community-developed core Python package and an ecosystem of tools and resources for astronomy \citep{astropy2022}, {\it NumPy}\footnote{https://numpy.org/doc/stable/index.html} \citep{harris2020a}, {\it SciPy}\footnote{https://scipy.org/} \citep{virtanen2020} scientific libraries used for scientific computing, {\it specutils}\footnote{https://specutils.readthedocs.io/en/stable/} \citep{specutils2019}, {\it extinction}\footnote{https://github.com/sncosmo/extinction} \citep{barbary2021}, and {\it pyphot}\footnote{https://mfouesneau.github.io/pyphot/} \citep{fouesneau2025}. We acknowledge with thanks the variable star observations from the AAVSO International Database\footnote{https://www.aavso.org/} contributed by observers worldwide and used in this research.

\section*{Data Availability}
The spectroscopic data and LCOGT photometric data underlying this article will be shared on reasonable request to the corresponding author. The other photometric data used in this work are from public surveys.



\bibliography{references} 



\appendix
\section{Post-Dimming Optical Spectrum of ASASSN-24fw}
\label{appendix:a}
\begin{figure*}
\centering
\includegraphics[width=0.95\textwidth]{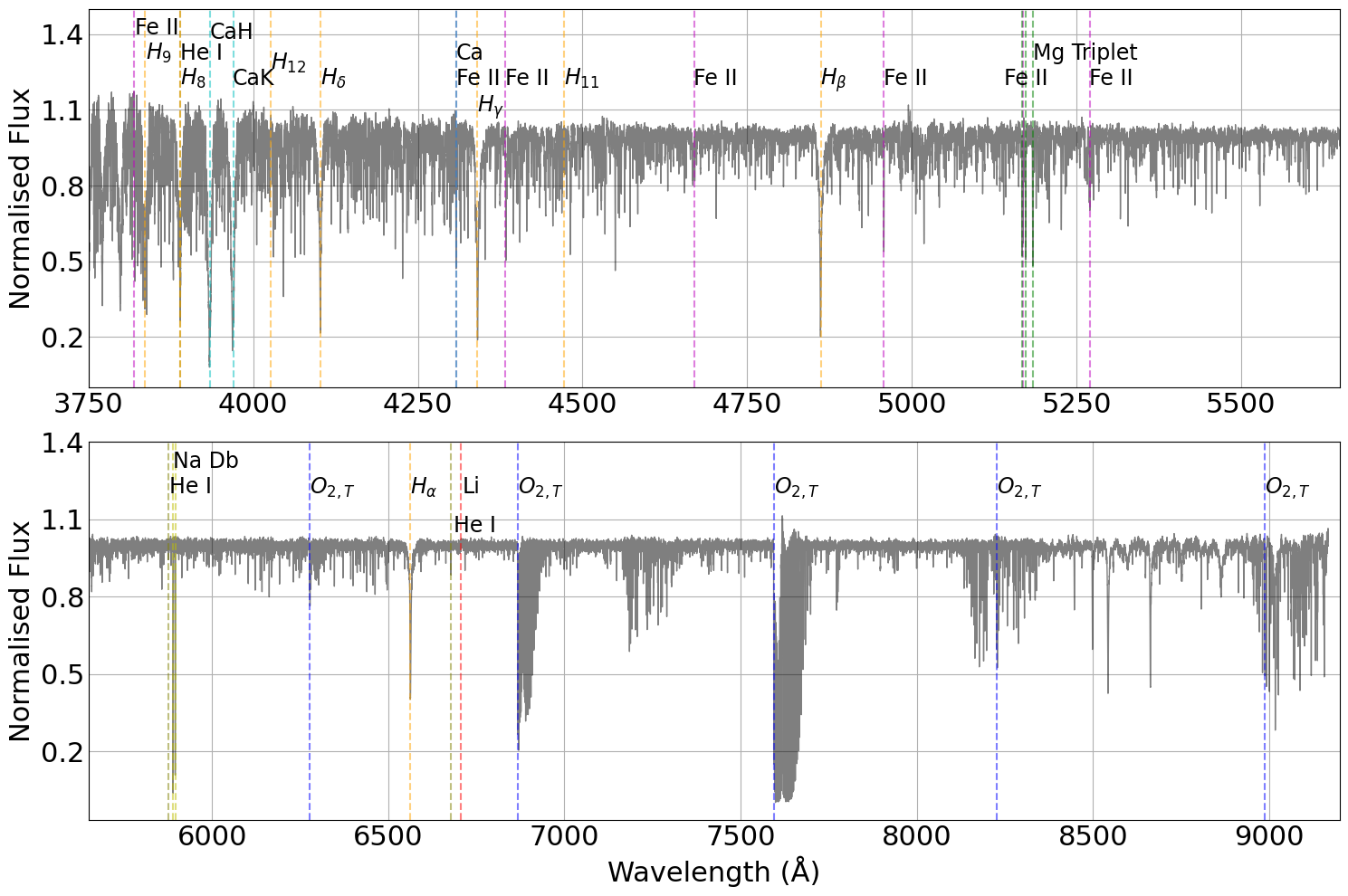}
\caption{The post-dimming optical spectrum of ASASSN-24fw obtained with MIKE on the Magellan Telescope.}
\label{appendixfig:a}
\end{figure*}

We observed ASASSN-24fw on 13 October 2025 using the MIKE Spectrograph on the Magellan Telescope.
Two exposures of 600\,s each were taken at a resolving power of $R \sim 57{,}000$.
The stacked, high-S/N spectrum is shown in Figure~\ref{appendixfig:a}, with line identifications marked in the same manner as in Figure~\ref{fig:lines}. The spectrum displays deep absorption in all prominent Balmer-series lines. 
The Ca\,\textsc{ii} H\&K lines and the Na\,\textsc{i} D doublet also exhibit strong absorption.
Most notably, H$\alpha$ appears in deep absorption, and the Li\,\textsc{i} 6708\,Å feature is negligible.
The absence of detectable lithium and the lack of emission features indicate that ASASSN-24fw is not a young stellar object (YSO) but instead an older main-sequence star.



\bsp	
\label{lastpage}
\end{document}